\documentclass[preprint,prd,showpacs,showkeys,eqsecnum,nofootinbib,unsortedaddress]{revtex4}

\usepackage{amsmath, amssymb, amsfonts, epsfig}
\usepackage{graphicx}

\begin{document}

\title{Phenomenology of a Light Cold Dark Matter Two-Singlet Model}
\author{Abdessamad Abada$^{\mathrm{a,b}}$}
\email{a.abada@uaeu.ac.ae}
\author{Salah Nasri$^{\mathrm{b}}$}
\email{snasri@uaeu.ac.ae}
\affiliation{$^{\mathrm{a}}$Laboratoire de Physique des Particules\ et Physique
Statistique, Ecole Normale Sup\'{e}rieure, BP 92 Vieux Kouba, 16050 Alger,
Algeria\\
$^{\mathrm{b}}$Physics Department, United Arab Emirates University, POB
17551, Al Ain, United Arab Emirates}

\begin{abstract}
We study the implications of phenomenological processes on a two-singlet
extension of the Standard Model we introduced in a previous work to describe
light cold dark matter. We look into the rare decays of $\Upsilon $ and $B$
mesons, most particularly the invisible channels, and study the decay
channels of the Higgs particle. Preferred regions of the parameter space are
indicated, together with others that are excluded. Comments in relation to
recent Higgs searches and finds at the LHC are made.
\end{abstract}

\date{\today }
\maketitle

\section{Introduction}

While still elusive, dark matter is believed to contribute about 23\% to the
energy budget of the Universe \cite{WMAP}. We know it should be massive,
stable on cosmic time scales and nonrelativistic when it decouples from the
thermal bath in order to be consistent with structure formation. Although
its mass and spin are not yet known, masses in the range of $5-10$ GeV seem
to be favored by the direct detection experiments CoGeNT \cite{CoGeNT}, DAMA 
\cite{DAMA} and CRESST II \cite{CRESST}.

A number of models have been proposed to try to explain the results of these
experiments \cite{models}. As it turns out, having a light dark-matter
candidate in supersymmetric theories is quite challenging. For instance, in
mSUGRA, the constraint from WMAP and the bound on the pseudo-scalar Higgs
mass from LEP give $m_{\chi _{1}^{0}}\geq 50\mathrm{GeV}$ \cite{msugra}.
Also, in the MSSM, a lightest supersymmetric particle with a mass around $10
\mathrm{GeV}$ and an elastic scattering cross-section off a nuclei as large
as $10^{-41}\mathrm{cm}^{2}$ is needed in order to fit the CoGeNT data,
which in turn requires a very large $\tan {\beta }$ and a relatively light
CP-odd Higgs. However, such a choice of parameters leads to a sizable
contribution to the branching ratios of some rare decays, which then
disfavors the scenario of light neutralinos in the context of the MSSM \cite
{MSSM} (see also \cite{Other}).

In a recent work \cite{abada-ghaffor-nasri}, we proposed a two-singlet
extension of the Standard Model as a simple model for light cold dark
matter. Both scalar fields were $\mathbb{Z}_{2}$-symmetric, with one
undergoing spontaneous symmetry breaking while the other remaining unbroken
to ensure stability of the dark-matter candidate. We studied the behavior of
the model, in particular the effects of the dark-matter relic-density
constraint and the restrictions from experimental direct detection. We
concluded that the model was capable of bearing a light dark-matter
weakly-interacting massive particle (WIMP) in mass regions where other
models may find difficulties. We should mention in passing that there are
scenarios with unstable light Higgs-like particles that have been previously
studied in certain extensions of the Standard Model, see for example \cite
{Ext}. There is also the possibility of having a light pseudo-scalar in the
NMSSM, see for example \cite{Gunion}.

The present work studies the effects and restrictions on the two-singlet
model coming from particle phenomenology. A limited selection of low-energy
processes has to be made, and we choose to look into the rare decays of $
\Upsilon $ and $B$ mesons. We limit ourselves to small dark-matter masses,
in the range $0.1-10$ GeV. We also study the implications of the model on
the decay channels of the Higgs particle and make quick comments in relation
to recent finds at the LHC.

The theory starts effectively with eight parameters \cite
{abada-ghaffor-nasri}. The spontaneous breaking of the electroweak and $
\mathbb{Z}_{2}$ symmetries introduces the two vacuum expectation values $v$
and $v_{1}$ respectively. The value of $v$ is fixed experimentally to be $
246 \mathrm{GeV}$ and we take $v_{1}=100\mathrm{GeV}$. Four of the
parameters are the three physical masses $m_{0}$ (dark-matter singlet $S_{0}$
), $m_{1}$ (the second singlet $S_{1}$) and $m_{h}$ (Higgs $h$), plus the
mixing angle $\theta $ between $h$ and $S_{1}$. We let $m_{1}$ vary in the
interval $0.1-10~\mathrm{GeV}$ and fix the Higgs mass to $m_{h}=125\mathrm{
GeV}$ \cite{ATLAS, CMS}, except in the part about the Higgs decays where we
let $m_{h}$ vary in the interval $100-200~\mathrm{GeV}$\footnote{
The exclusion mass range reported by the CMS and ATLAS Collaborations
applies to the SM Higgs and can be weakened or evaded in models where the
Higgs production and/or decay channels are suppressed \cite{Hewett}. We will
comment on this possibility within our model in the last section.}. For the
purpose of our discussions, it is sufficient to let $\theta $ vary in the
interval $1^{\mathrm{o}}-40^{\mathrm{o}}$. The last parameters are the two
physical mutual coupling constants $\lambda _{0}^{\left( 4\right) }$ (dark
matter -- Higgs) and $\eta _{01}^{\left( 4\right) }$ (dark matter -- $S_{1}$
particle). In fact, $\eta _{01}^{\left( 4\right) }$ is not free as it is the
smallest real and positive solution to the dark-matter relic density
constraint \cite{abada-ghaffor-nasri}, which is implemented systematically
throughout this work. Thus we are left with four parameters, namely, $m_{0}$
, $m_{1}$, $\theta $ and $\lambda _{0}^{(4)}$. To ensure applicability of
perturbation theory, the requirement $\eta _{01}^{\left( 4\right) }<1$ is
also imposed throughout, as well as a choice of rather small values for $
\lambda _{0}^{\left( 4\right) }$.

\section{Upsilon decays}

We start by looking at the constraints on the parameter space of the model
coming from the decay of the meson $\Upsilon $ in the state $nS$ ($n=1,3$)
into one photon $\gamma $ and one particle $S_{1}$. For $m_{1}\lesssim 8 
\mathrm{GeV}$, the branching ratio for this process is given by the
relation: 
\begin{equation}
\mathrm{Br}\left( \Upsilon _{nS}\rightarrow \gamma +S_{1}\right) =\frac{
G_{F}m_{b}^{2}\sin ^{2}\theta }{\sqrt{2}\pi \alpha }\;x_{n}\left( 1-\frac{
4\alpha _{s}}{3\pi }f(x_{n})\right) \mathrm{Br}^{(\mu )}\,\Theta \left(
m_{\Upsilon _{nS}}-m_{1}\right) .  \label{BrUpsilon-photon-S1}
\end{equation}
In this expression, $x_{n}\equiv \left( 1-m_{1}^{2}/m_{\Upsilon
_{ns}}^{2}\right) $ with $m_{\Upsilon _{1(3)S}}=$ $9.46(10.355)\mathrm{GeV}$
the mass of $\Upsilon _{1(3)S}$, the branching ratio $\mathrm{Br}^{(\mu
)}\equiv \mathrm{Br}\left( \Upsilon _{1(3)S}\rightarrow \mu ^{+}\mu
^{-}\right) =2.48(2.18)\times 10^{-2}$ \cite{Eidel}, $\alpha $ is the QED
coupling constant, $\alpha _{s}=0.184$ the QCD\ coupling constant at the
scale $m_{\Upsilon _{nS}}$, the quantity $G_{F}$ is the Fermi coupling
constant and $m_{b}$ the $b$ quark mass \cite{PDG}. The function $f(x)$
incorporates the effect of QCD radiative corrections given in \cite{QCD}.

However, a rough estimate of the lifetime of $S_{1}$ indicates that this
latter is likely to decay inside a typical particle detector, which means we
ought to take into account its most dominant decay products. We first have a
process by which $S_{1}$ decays into a pair of pions, with a decay rate
given by: 
\begin{eqnarray}
\Gamma \left( S_{1}\rightarrow \pi \pi \right) &=&\frac{G_{F}m_{1}}{4\sqrt{2}
\pi }\sin ^{2}\theta \left[ \frac{m_{1}^{2}}{27}\left( 1+\frac{11m_{\pi
}^{2} }{2m_{1}^{2}}\right) ^{2}\right.  \notag \\
&&\times \left( 1-\frac{4m_{\pi }^{2}}{m_{1}^{2}}\right) ^{\frac{1}{2}
}\Theta \lbrack \left( m_{1}-2m_{\pi }\right) \left( 2m_{K}-m_{1}\right) ] 
\notag \\
&&\left. +3\left( M_{u}^{2}+M_{d}^{2}\right) \left( 1-\frac{4m_{\pi }^{2}}{
m_{1}^{2}}\right) ^{\frac{3}{2}}\Theta \left( m_{1}-2m_{K}\right) \right] .
\label{decay-rate-S1-pipi}
\end{eqnarray}
Here, $m_{\pi }$ is the pion mass and $m_{K}$ the kaon mass. Also, chiral
perturbation theory is used below the kaon pair production threshold, and
the MIT bag model above, with the dressed $u$ and $d$ quark masses $
M_{u}=M_{d}=0.05\mathrm{GeV}$. Note that this rate includes all pions,
charged and neutral. Above the $2m_{K}$ threshold, there is the production
of both a pair of kaons and $\eta $ particles. The decay rate for $K$
production is: 
\begin{equation}
\Gamma \left( S_{1}\rightarrow KK\right) =\frac{9}{13}\frac{
3G_{F}M_{s}^{2}m_{1}}{4\sqrt{2}\pi }\sin ^{2}\theta \left( 1-\frac{
4m_{K}^{2} }{m_{1}^{2}}\right) ^{\frac{3}{2}}\Theta \left(
m_{1}-2m_{K}\right) .  \label{decay-rate-S1-KK}
\end{equation}
In the above rate, $M_{s}=0.45\mathrm{GeV}$ is the $s$ quark bag-mass \cite
{Hunters, MCkeen}. For $\eta $ production, replace $m_{K}$ by $m_{\eta }$
and $\frac{9}{13}$ by $\frac{4}{13}$.

The particle $S_{1}$ decays also into $c$ and $b$ quarks (mainly $c$).
Including the radiative QCD corrections, the corresponding decay rates are
given by: 
\begin{equation}
\Gamma (S_{1}\rightarrow q\bar{q})=\frac{3G_{F}\bar{m}_{q}^{2}m_{1}}{4\sqrt{
2 }\pi }\sin ^{2}\theta \left( 1-\frac{4\bar{m}_{q}^{2}}{m_{h}^{2}}\right)
^{ \frac{3}{2}}\left( 1+5.67\frac{\bar{\alpha}_{s}}{\pi }\right) \Theta
\left( m_{1}-2\bar{m}_{q}\right) .  \label{decay-rate-S1-qq}
\end{equation}
The dressed quark mass $\bar{m}_{q}\equiv m_{q}(m_{1})$ and the running
strong coupling constant $\bar{\alpha}_{s}\equiv \alpha _{s}(m_{1})$ are
defined at the energy scale $m_{1}$ \cite{Djouadi}. There is also a decay
into a pair of gluons, with the rate: 
\begin{equation}
\Gamma \left( S_{1}\rightarrow gg\right) =\frac{G_{F}m_{1}^{3}\sin
^{2}\theta }{12\sqrt{2}\pi }\left( \frac{\alpha _{s}^{\prime }}{\pi }\right)
^{2}\left[ 6-2\left( 1-\frac{4m_{\pi }^{2}}{m_{1}^{2}}\right) ^{\frac{3}{2}
}-\left( 1-\frac{4m_{K}^{2}}{m_{1}^{2}}\right) ^{\frac{3}{2}}\right] \Theta
\left( m_{1}-2m_{\pi }\right) .  \label{decay-rate-S1-gg}
\end{equation}
Here, $\alpha _{s}^{\prime }=0.47$ is the QCD coupling constant at the
bag-model scale.

We then have the decay of $S_{1}$ into leptons, the corresponding rate given
by: 
\begin{equation}
\Gamma \left( S_{1}\rightarrow \ell ^{+}\ell ^{-}\right) =\frac{G_{F}m_{\ell
}^{2}m_{1}}{4\sqrt{2}\pi }\sin ^{2}\theta \left( 1-\frac{4m_{\ell }^{2}}{
m_{1}^{2}}\right) ^{\frac{3}{2}}\Theta \left( m_{1}-2m_{\ell }\right) ,
\label{decay-rate-S1-ff}
\end{equation}
where $m_{\ell }$ is the lepton mass. Finally, $S_{1}$ can decay into a pair
of dark matter particles, with a decay rate: 
\begin{equation}
\Gamma \left( S_{1}\rightarrow S_{0}S_{0}\right) =\frac{\left( \eta
_{01}^{(3)}\right) ^{2}}{32\pi m_{1}}\sqrt{1-\frac{4m_{0}^{2}}{m_{1}^{2}}}
\Theta \left( m_{1}-2m_{0}\right) .  \label{decay-rate-S1-S0S0}
\end{equation}
The coupling constant $\eta _{01}^{(3)}$ is given in \cite
{abada-ghaffor-nasri}. The branching ratio for $\Upsilon _{nS}$ decaying via 
$S_{1}$ into a photon plus $X$, where $X$ represents any kinematically
allowed final state, will be: 
\begin{equation}
\mathrm{Br}\left( \Upsilon _{nS}\rightarrow \gamma +X\right) =\mathrm{Br}
\left( \Upsilon _{nS}\rightarrow \gamma +S_{1}\right) \times \mathrm{Br}
\left( S_{1}\rightarrow X\right) .  \label{BrUpsilon1s-photon-X}
\end{equation}
In particular, $X\equiv S_{0}S_{0}$ corresponds to a decay into invisible
particles.

\begin{figure}[htb]
\centering
\includegraphics[width=6.52in,height=3.7377in]{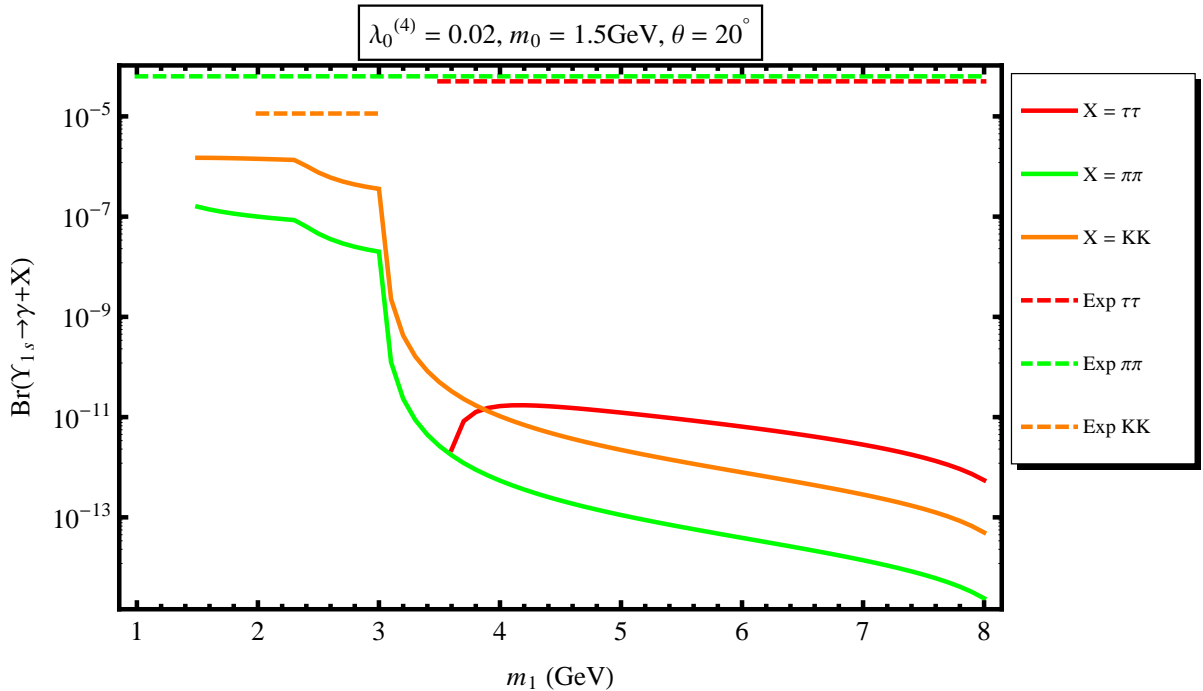}
\caption{Typical branching ratios of $\Upsilon _{1S}$ decaying into $\protect
\tau $'s, charged pions and charged kaons as functions of $m_{1}$. The
corresponding experimental upper bounds are shown.}
\label{BrUps1s_viaS1_lambda04-002_m0-15_theta-20}
\end{figure}

\begin{figure}[htb]
\centering
\includegraphics[width=6.52in,height=3.7377in]{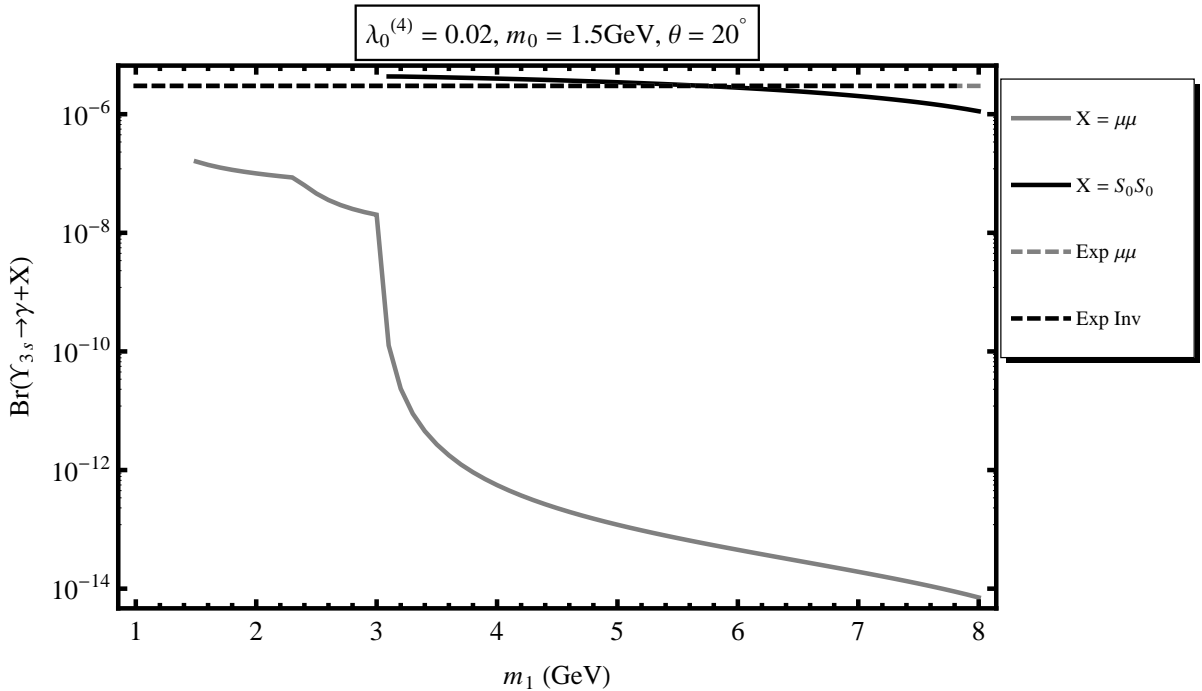}
\caption{Typical branching ratios of $\Upsilon _{3S}$ decaying into muons
and dark matter as functions of $m_{1}$. The corresponding experimental
upper bounds are shown.}
\label{brups3s_vias1_lambda04-002_m0-15_theta-20}
\end{figure}

The best available experimental upper bounds on $1S$--state branching ratios
are: (i) $\mathrm{Br}\left( \Upsilon _{1S}\rightarrow \gamma +\tau \tau
\right) <5\times 10^{-5}$ for $3.5\mathrm{GeV}<m_{1}<9.2\mathrm{GeV}$ \cite
{upstau}; (ii) $\mathrm{Br}\left( \Upsilon _{1S}\rightarrow \gamma +\pi
^{+}\pi ^{-}\right) <6.3\times 10^{-5}$ for $1\mathrm{GeV}<m_{1}$ \cite
{upspp}; (iii) $\mathrm{Br}\left( \Upsilon _{1S}\rightarrow \gamma
+K^{+}K^{-}\right) <1.14\times 10^{-5}$ for $2\mathrm{GeV}<m_{1}<3\mathrm{GeV
}$ \cite{upskk}. Figure \ref{BrUps1s_viaS1_lambda04-002_m0-15_theta-20}
displays the corresponding branching ratios of $\Upsilon _{1S}$ decays via $
S_{1}$ as functions of $m_{1}$, together with these upper bounds. Also, the
best available experimental upper bounds on $\Upsilon _{3S}$ branching
ratios are: (i) $\mathrm{Br}\left( \Upsilon _{3S}\rightarrow \gamma +\mu \mu
\right) <3\times 10^{-6}$ for $1\mathrm{GeV}<m_{1}<10\mathrm{GeV}$; (ii) $
\mathrm{Br}\left( \Upsilon _{3S}\rightarrow \gamma +\mathrm{Invisible}
\right) <3\times 10^{-6}$ for $1\mathrm{GeV}<m_{1}<7.8\mathrm{GeV}$ \cite
{3smu}. Typical corresponding branching ratios are shown in figure \ref
{brups3s_vias1_lambda04-002_m0-15_theta-20}.

A systematic scan of the parameter space indicates that the main effect of
the Higgs-dark-matter coupling constant $\lambda _{0}^{(4)}$ and the
dark-matter mass $m_{0}$ is to exclude, via the relic density and
perturbativity constraints, regions of applicability of the model. This is
shows in figures \ref{BrUps1s_viaS1_lambda04-002_m0-15_theta-20} and \ref
{brups3s_vias1_lambda04-002_m0-15_theta-20} where the region $m_{1}\lesssim
1.4\mathrm{GeV}$ is excluded. Otherwise, these two parameters have little
effect on the shapes of the branching ratios themselves. The onset of the $
S_{0}S_{0}$ channel for $m_{1}\geq 2m_{0}$ abates sharply the other channels
and this one becomes dominant by far. The effect of the mixing angle $\theta 
$ is to enhance all branching ratios as it increases, due to the factor $
\sin ^{2}\theta $. The dark matter decay channel reaches the invisible upper
bound already for $\theta \simeq 15^{\mathrm{o}}$, for fairly small $m_{0}$,
say 0.5GeV. The other channels find it hard to get to their respective
experimental upper bounds, even for large values of $\theta $.

\section{$B$ meson decays}

Next we look at the flavor changing process in which the meson $B^{+}$
decays into a $K^{+}$ plus invisibles. The corresponding Standard-Model mode
is a decay into $K^{+}$ and a pair of neutrinos, with a branching ratio $
\mathrm{Br}^{\mathrm{SM}}\left( B^{+}\rightarrow K^{+}+\nu \bar{\nu}\right)
\simeq 4.7\times 10^{-6}$ \cite{BKSM}. The experimental upper bound is $
\mathrm{Br}^{\mathrm{Exp}}\left( B^{+}\rightarrow K^{+}+\mathrm{Inv}\right)
\simeq 14\times 10^{-6}$ \cite{BK}. As in the $\Upsilon $ decays, the most
prominent $B$ invisible decay in this model is into $S_{0}S_{0}$ via $S_{1}$
. The process $B^{+}\rightarrow K^{+}+S_{1}$ has a the branching ratio: 
\begin{eqnarray}
\mathrm{Br}\left( B^{+}\rightarrow K^{+}+S_{1}\right) &=&\frac{9\sqrt{2}\tau
_{B}G_{F}^{3}m_{t}^{4}m_{b}^{2}m_{+}^{2}m_{-}^{2}}{1024\pi
^{5}m_{B}^{3}\left( m_{b}-m_{s}\right) ^{2}}\left\vert V_{tb}V_{ts}^{\ast
}\right\vert ^{2}f_{0}^{2}\left( m_{1}^{2}\right)  \notag \\
&&\times \sqrt{\left( m_{+}^{2}-m_{1}^{2}\right) \left(
m_{-}^{2}-m_{1}^{2}\right) }\;\sin ^{2}\theta \;\Theta \left(
m_{-}-m_{1}\right) .  \label{Br-Bplus-S1}
\end{eqnarray}
Here $m_{\pm }=m_{B}\pm m_{K}$ where $m_{B}$ is the $B^{+}$ mass, $\tau _{B}$
its lifetime, and$V_{tb}$ and $V_{ts}$ are flavor changing CKM coefficients.
The function $f_{0}\left( s\right) $ is given by the relation: 
\begin{equation}
f_{0}\left( s\right) =0.33\exp \left[ \frac{0.63s}{m_{B}^{2}}-\frac{
0.095s^{2}}{m_{B}^{4}}+\frac{0.591s^{3}}{m_{B}^{6}}\right] .
\label{f_0-in-Bplus-decay}
\end{equation}
The different $S_{1}$ decay modes are given in (\ref{decay-rate-S1-pipi}) -
( \ref{decay-rate-S1-S0S0}) above. The branching ratio of $B^{+}$ decaying
into $K^{+}+S_{0}S_{0}$ via the production and propagation of an
intermediary $S_{1}$ will be: 
\begin{equation}
\mathrm{Br}^{(S_{1})}\left( B^{+}\rightarrow K^{+}+S_{0}S_{0}\right) = 
\mathrm{Br}\left( B^{+}\rightarrow K^{+}+S_{1}\right) \times \mathrm{Br}
\left( S_{1}\rightarrow S_{0}S_{0}\right) .  \label{BrBplus-K-X-via-S1}
\end{equation}

\begin{figure}[htb]
\centering
\includegraphics[width=6.52in,height=3.8449in]
{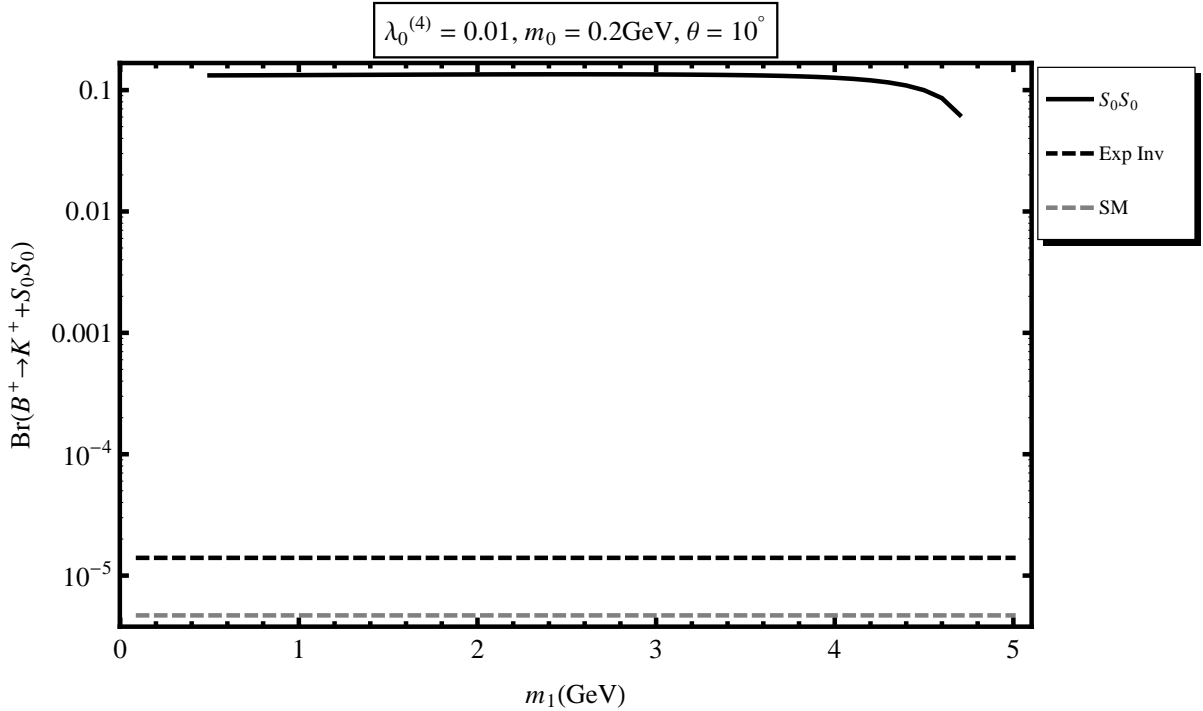}
\caption{Typical branching ratio of $B^{+}$ decaying into dark matter via $
S_{1}$ as a function of $m_{1}$. The SM and experimental bounds are shown.}
\label{Bp_bran-ratio_lambda04-001_m0-02_theta-10}
\end{figure}

Figure \ref{Bp_bran-ratio_lambda04-001_m0-02_theta-10} displays a typical
behavior of $\mathrm{Br}^{(S_{1})}\left( B^{+}\rightarrow
K^{+}+S_{0}S_{0}\right) $ as a function of $m_{1}$. The branching ratio is
well above the experimental upper bound, and $\theta $ as small as $1^{ 
\mathrm{o}}$ will not help with this, no matter what the values for $\lambda
_{0}^{(4)}$ and $m_{0}$ are. So, for $m_{1}\lesssim 4.8\mathrm{GeV}$, this
process excludes the two-singlet model for $m_{0}<m_{1}/2$. For $
m_{1}\gtrsim 4.8\mathrm{GeV}$ or $m_{0}\geq m_{1}/2$, the decay does not
occur, so no constraints on the model from this process.

Another process involving $B$ mesons is the decay of $B_{s}$ into
predominately a pair of muons. The Standard Model branching ratio for this
process is $\mathrm{Br}^{\mathrm{SM}}\left( B_{s}\rightarrow \mu ^{+}\mu
^{-}\right) =(3.2\pm 0.2)\times 10^{-9}$ \cite{BSSM}, and the experimental
upper bound is $\mathrm{Br}^{\mathrm{Exp}}\left( B_{s}\rightarrow \mu
^{+}\mu ^{-}\right) <1.08\times 10^{-8}$ \cite{BSEXP}. In the present model,
two additional decay diagrams occur, both via intermediary $S_{1}$, yielding
together the branching ratio: 
\begin{equation}
\mathrm{Br}^{(S_{1})}(B_{s}\rightarrow \mu ^{+}\mu ^{-})=\frac{9\tau
_{B_{s}}G_{F}^{4}f_{B_{s}}^{2}m_{B_{s}}^{5}}{2048\pi ^{5}}m_{\mu
}^{2}m_{t}^{4}\left\vert V_{tb}V_{ts}^{\ast }\right\vert ^{2}\frac{\left(
1-4m_{\mu }^{2}/m_{B_{s}}^{2}\right) ^{3/2}}{\left(
m_{B_{s}}^{2}-m_{1}^{2}\right) ^{2}+ m^2_1\Gamma _{1}^{2}}\sin ^{4}\theta .
\label{B_s branching via S_1}
\end{equation}
In this relation, $\tau _{B_{s}}$ is the $B_{s}$ life-time, $m_{B_{s}}= 5.37 
\mathrm{GeV}$ its mass, and $f_{B_{s}}$\ a form factor that we take equal to 
$0.21$GeV. The quantity $\Gamma _{1}$ is the total width of the particle $
S_{1}$  \cite{abada-ghaffor-nasri}.

\begin{figure}[htb]
\centering
\includegraphics[width=6.52in,height=3.8792in]
{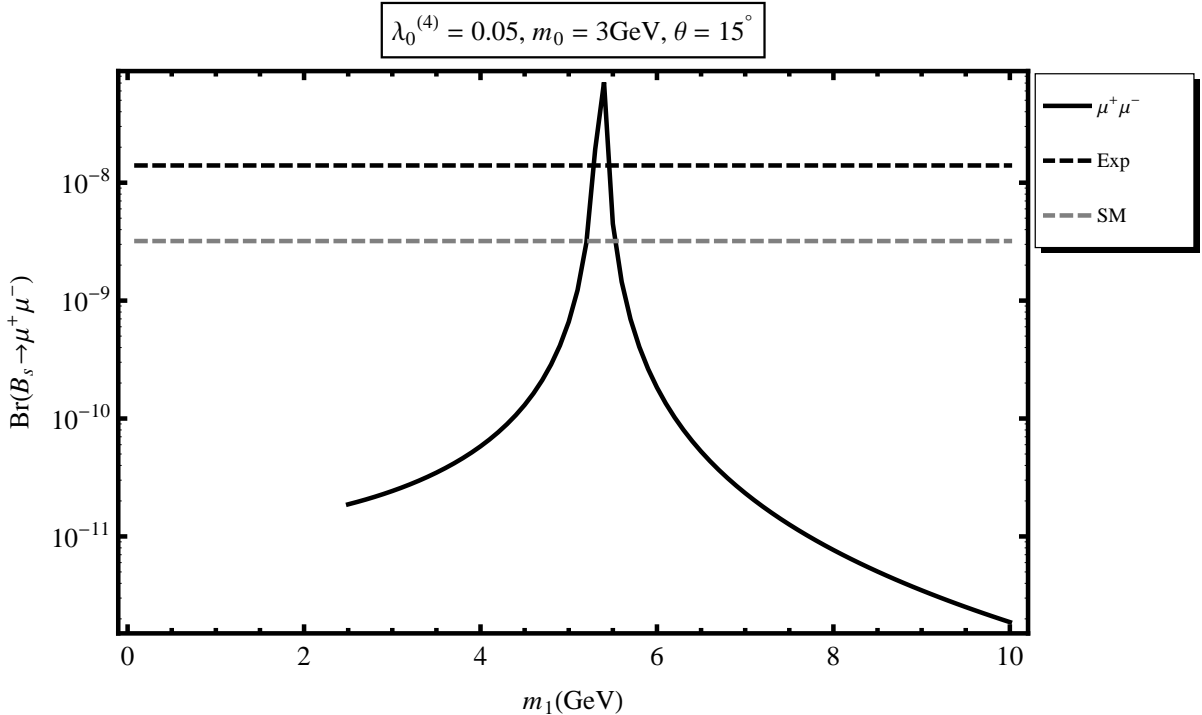}
\caption{Typical behavior of $\mathrm{Br}^{(S_{1})}(B_{s}\rightarrow \protect
\mu ^{+}\protect\mu ^{-})$ as a function of $m_{1}$, together with the SM
and experimental bounds.}
\label{Bs_branch-ratio_lambda04-005_m0-3_theta-15}
\end{figure}

A typical behavior of $\mathrm{Br}^{(S_{1})}(B_{s}\rightarrow \mu ^{+}\mu
^{-})$ as a function of $m_{1}$ is shown in figure \ref
{Bs_branch-ratio_lambda04-005_m0-3_theta-15}. The peak is at $m_{B_{s}}$.
All three parameters $\lambda _{0}^{(4)}$, $m_{0}$ and $\theta $ combine in
the relic density constraint to exclude regions of applicability of the
model. For example, for the values of figure \ref
{Bs_branch-ratio_lambda04-005_m0-3_theta-15}, the region $m_{1}<2.2\mathrm{\
GeV}$ is excluded. However, a systematic scan of the parameter space shows
that outside the relic density constraint, $\lambda _{0}^{(4)}$ has no
significant direct effect on the shape of $\mathrm{Br}^{(S_{1})}(B_{s}
\rightarrow \mu ^{+}\mu ^{-})$. As $m_{0}$ increases, it sharpens the peak
of the curve while pushing it up. This works until about 2.7GeV, beyond
which $m_{0}$ ceases to have any significant direct effect. Increasing $
\theta $ enhances the values of the branching ratio without affecting the
width.

\begin{figure}[htb]
\centering
\includegraphics[width=3.5653in,height=3.9101in]
{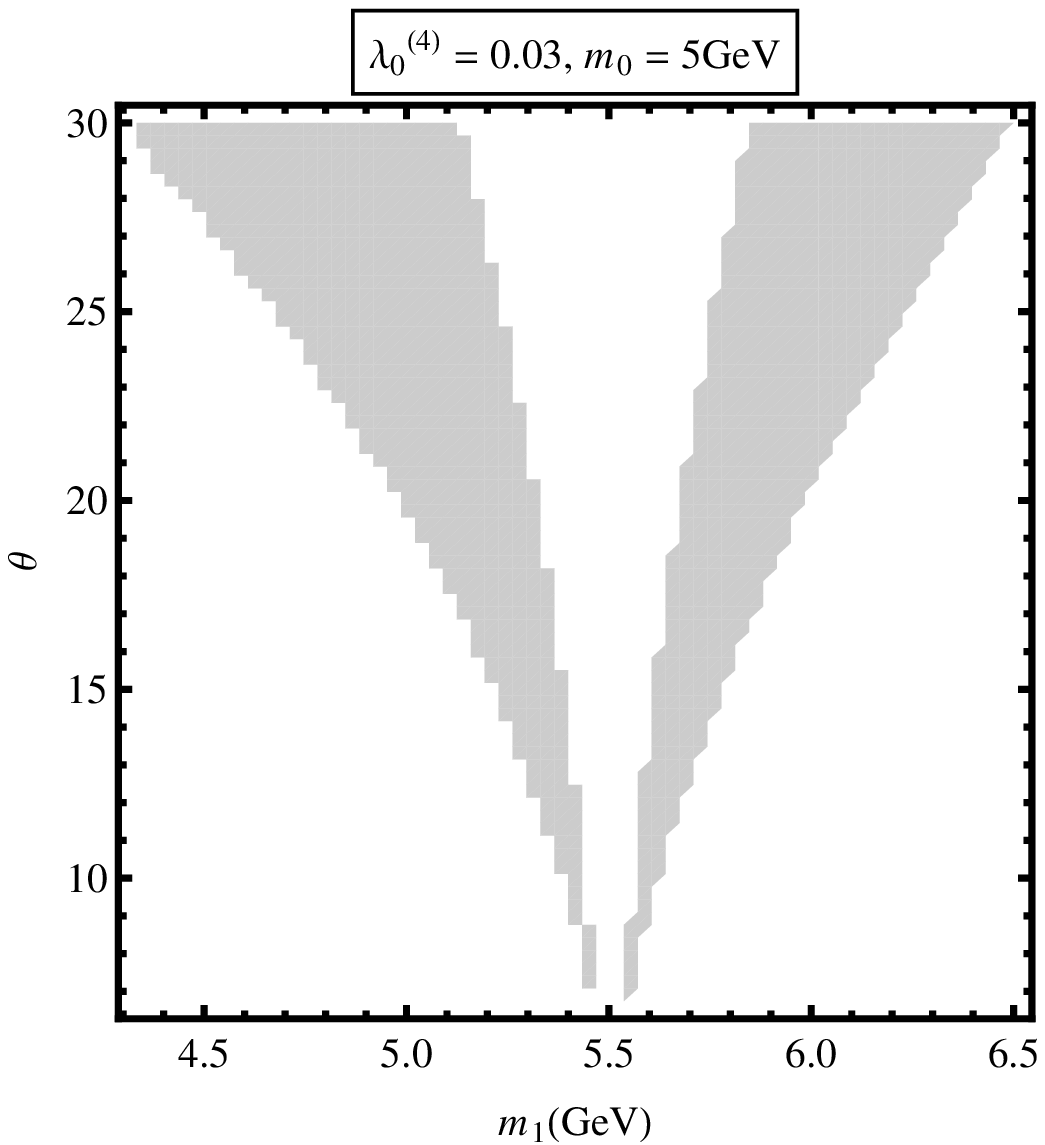}
\caption{A density plot of $B^{\mathrm{SM}}+\mathrm{Br}^{(S_{1})}$ squeezed
between $B^{\mathrm{SM}}+ 5\protect\sigma $ from below and $B^{\mathrm{Exp}}$
from above.}
\label{Bs_branch-ratio_lambda04-003_m0-5}
\end{figure}

Also, for all the range of $m_{1}$, all of $\mathrm{Br}^{(S_{1})}+\mathrm{Br}
^{\mathrm{SM}}$ stays below $\mathrm{Br}^{\mathrm{Exp}}$ as long as $\theta
<10^{\mathrm{o}}$. As $\theta $ increases beyond this value, the peak region
pushes up increasingly above $\mathrm{Br}^{\mathrm{Exp}}$ and thus gets
excluded. Hoping for a clear signal if any, figure \ref
{Bs_branch-ratio_lambda04-003_m0-5} displays a density plot showing $\mathrm{
\ Br}^{\mathrm{SM}}$+$\mathrm{Br}^{(S_{1})}(B_{s}\rightarrow \mu ^{+}\mu
^{-})$ in the plane $\left( m_{1},\theta \right) $, squeezed between $
\mathrm{Br}^{ \mathrm{SM}}+5\sigma $ from below and $\mathrm{Br}^{\mathrm{Exp
}}$ from above. The behavior we see in this figure is generic across the
ranges of $m_{0}$ and $\lambda _{0}^{(4)}$: the V-shape structure in gray
developing from $m_{1}=m_{B_{s}}$ is the allowed region. The white region in
the middle is due to $\mathrm{Br}^{\mathrm{Exp}}$, and the white region
outside to $\mathrm{Br}^{\mathrm{SM}}+5\sigma $. It can happen that some of
the gray V is eaten up by the relic-density constraint and perturbativity
requirement for larger values of $\lambda _{0}^{(4)}$.

From this process, there is probably one element to retain if we want the
model to contribute a distinct signal to $B_{s}\rightarrow \mu ^{+}\mu ^{-}$
for the range of $m_{0}$ chosen, and that is to restrict $4\mathrm{GeV}
\lesssim m_{1}\lesssim 6.5\mathrm{GeV}$. No additional constraint on $m_{0}$
is necessary while keeping $\lambda _{0}^{(4)}\lesssim 0.1$ to avoid
systematic exclusion from direct detection is safe.

\section{Higgs decays}

We finally examine the implications of the model on the Higgs different
decay modes. In this part of the work, we allow the Higgs mass $m_{h}$ to
vary in the interval $100\mathrm{GeV}-200\mathrm{GeV}$. First, $h$ can decay
into a pair of leptons $\ell $, predominantly $\tau $'s. The corresponding
decay rate $\Gamma \left( h\rightarrow \ell ^{+}\ell ^{-}\right) $ is given
by the relation (\ref{decay-rate-S1-ff}) where we replace $m_{1}$ by $m_{h}$
. It can also decay into a pair of quarks $q$, mainly into $b$'s and, to a
lesser degree, into $c$'s. Here too the decay rate\ $\Gamma (h\rightarrow q
\bar{q})$ is given in (\ref{decay-rate-S1-qq}) with the replacement $m_{h}$
instead of $m_{1}$. Then the Higgs can decay into a pair of gluons.
Including the next-to-next-to-leading QCD radiative corrections, the
corresponding decay rate can write like this: 
\begin{eqnarray}
\Gamma \left( h\rightarrow gg\right)  &=&\frac{G_{F}m_{h}^{3}}{4\sqrt{2}\pi }
\left\vert \sum_{q}\frac{m_{q}^{2}}{m_{h}^{2}}\int_{0}^{1}dx\int_{0}^{1-x}dy
\frac{1-4xy}{\frac{m_{q}^{2}}{m_{h}^{2}}-xy}\right\vert ^{2}  \notag \\
&&\times \left( \frac{\bar{\alpha}_{s}}{\pi }\right) ^{2}\left[ 1+\frac{215}{
12}\frac{\bar{\alpha}_{s}}{\pi }+\frac{\bar{\alpha}_{s}^{2}}{\pi ^{2}}\left(
156.8-5.7\log {\frac{m_{t}^{2}}{m_{h}^{2}}}\right) \right] \cos ^{2}\theta ,
\label{higgs-gg-decay}
\end{eqnarray}
where the sum is over all quark flavors $q$. A systematic study of the
double integral above shows that, with $m_{h}$ in the range 100GeV --
200GeV, the $t$ quark dominates in the sum over $q$, with non-negligible
contributions from the $c$ and $b$ quarks.

For $m_{h}$ smaller than the $W$ or $Z$ pair-production threshold, the Higgs
can decay into a pair of one real and one virtual gauge bosons, with rates
given by: 
\begin{equation}
\Gamma \left( h\rightarrow VV^{\ast }\right) =\frac{3G_{F}^{2}m_{V}^{4}m_{h} 
}{16\pi ^{3}}\cos ^{2}\theta \,A_{V}\,R\left( \frac{m_{V}^{2}}{{m_{h}^{2}}}
\right) \Theta \left[ \left( m_{h}-m_{V}\right) \left( 2m_{V}-m_{h}\right) 
\right] .  \label{higgs-VVv-decay}
\end{equation}
In this expression, $m_{V}$ is the mass of the gauge boson $V$, the factor $
A_{V}=1$ for $W$ and $\left( \frac{7}{12}-\frac{10}{9}\sin ^{2}{\theta _{w}}
+ \frac{40}{9}\sin ^{4}{\theta _{w}}\right) $ for $Z$ with $\theta _{w}$ the
Weinberg angle, and we have: 
\begin{eqnarray}
R(x) &=&\frac{3(1-8x+20x^{2})}{\sqrt{4x-1}}\arccos {\left( \frac{3x-1}{
2x^{3/2}}\right) }  \notag \\
&&-\frac{1-x}{2x}\left( 2-13x+47x^{2}\right) -\frac{3}{2}\left(
1-6x+4x^{2}\right) \log {x.}  \label{R-in-h--WWv}
\end{eqnarray}
For a heavier Higgs particle, the decay rates into a $V$ pair is given by: 
\begin{equation}
\Gamma \left( h\rightarrow VV\right) =\hspace{-2pt}\frac{G_{F}m_{V}^{4}\cos
^{2}\theta }{\sqrt{2}\pi m_{h}}B_{V}\left( 1-\frac{4m_{V}^{2}}{m_{h}^{2}}
\right) ^{\frac{1}{2}}\left[ 1+\frac{\left( m_{h}^{2}-2m_{V}^{2}\right) ^{2} 
}{8m_{V}^{4}}\right] \Theta \left( m_{h}-2m_{V}\right) ,
\label{higgs-VV-decay}
\end{equation}
with $B_{V}=1$ for $W$ and $\frac{1}{2}$ for $Z$.

While all these decay modes already exist within the Standard Model, the
two-singlet extension introduces two additional (invisible) modes, namely a
decay into a pair of $S_{0}$'s and a pair of $S_{1}$'s. The corresponding
decay rates are: 
\begin{equation}
\Gamma \left( h\rightarrow S_{i}S_{i}\right) =\frac{\lambda _{i}^{2}}{32\pi
m_{h}}\left( 1-\frac{4m_{i}^{2}}{m_{h}^{2}}\right) ^{\frac{1}{2}}\Theta
\left( m_{h}-2m_{i}\right) ,  \label{higgs-SiSi-decay}
\end{equation}
where $\lambda _i=\lambda _{0\left( 2\right) }^{(3)}$ for $S_{0\left(
1\right) }$ are coupling constants given in \cite{abada-ghaffor-nasri},
functions of the parameters of the theory. The total decay rate $\Gamma
\left( h\right) $ of the Higgs particle is the sum of these partial rates.
The branching ratio corresponding to a particular decay will be $\mathrm{Br}
\left( h\rightarrow X\right) =\Gamma \left( h\rightarrow X\right) /\Gamma
\left( h\right) $.

\begin{figure}[htb]
\centering
\includegraphics[width=5.929in,height=3.4307in]
{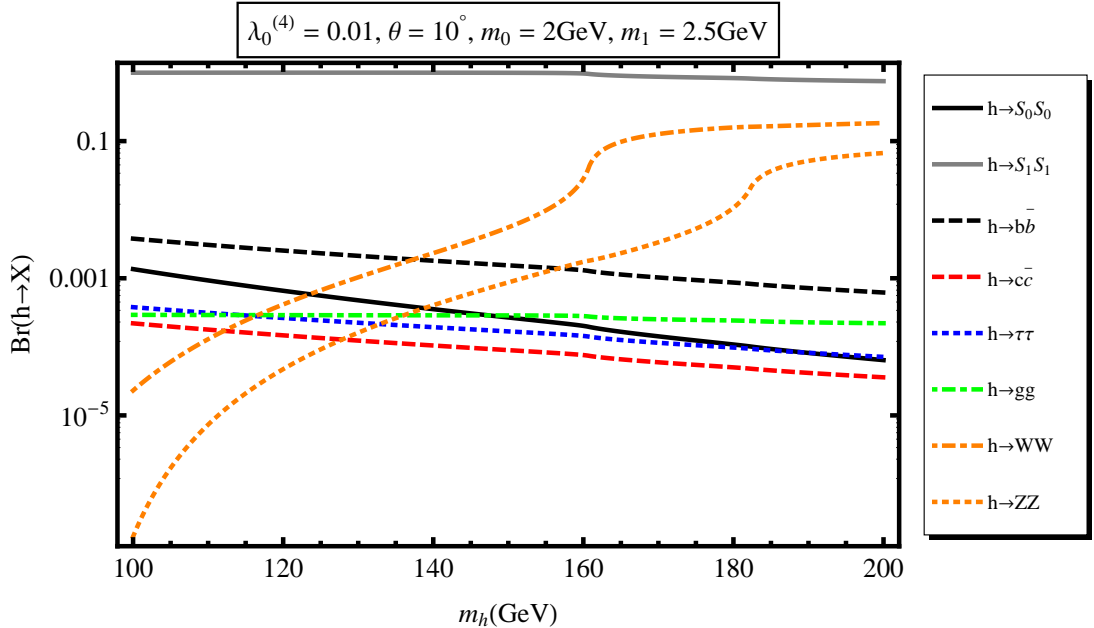}
\caption{Branching ratios for Higgs decays. Very small dark-matter Higgs
coupling.}
\label{log-higgs_lambda04-001_theta-10_m0-2_m1-25_v2}
\end{figure}

Typical behaviors of the most prominent branching ratios are displayed in
figure \ref{log-higgs_lambda04-001_theta-10_m0-2_m1-25_v2}. A systematic
study shows that for all ranges of the parameters, the Higgs decays
dominantly into invisible. The production of fermions and gluons is
comparatively marginal, whereas that of $W$ and $Z$ pairs takes relative
importance towards and above the corresponding thresholds, and more
significantly at larger values of the mixing angle $\theta $.

\begin{figure}[htb]
\centering
\includegraphics[width=5.8802in,height=3.4024in]
{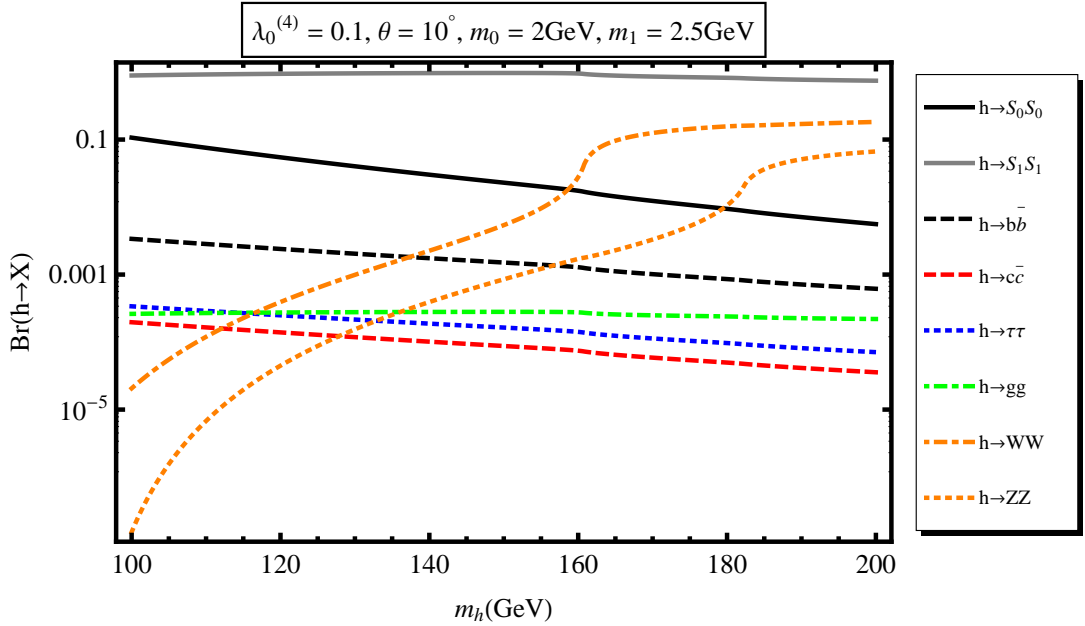}
\caption{Branching ratios for Higgs decays. Small dark-matter Higgs
coupling. }
\label{log-higgs_lambda04-01_theta-10_m0-2_m1-25_v2}
\end{figure}

\begin{figure}[htb]
\centering
\includegraphics[width=5.8802in,height=3.4718in]
{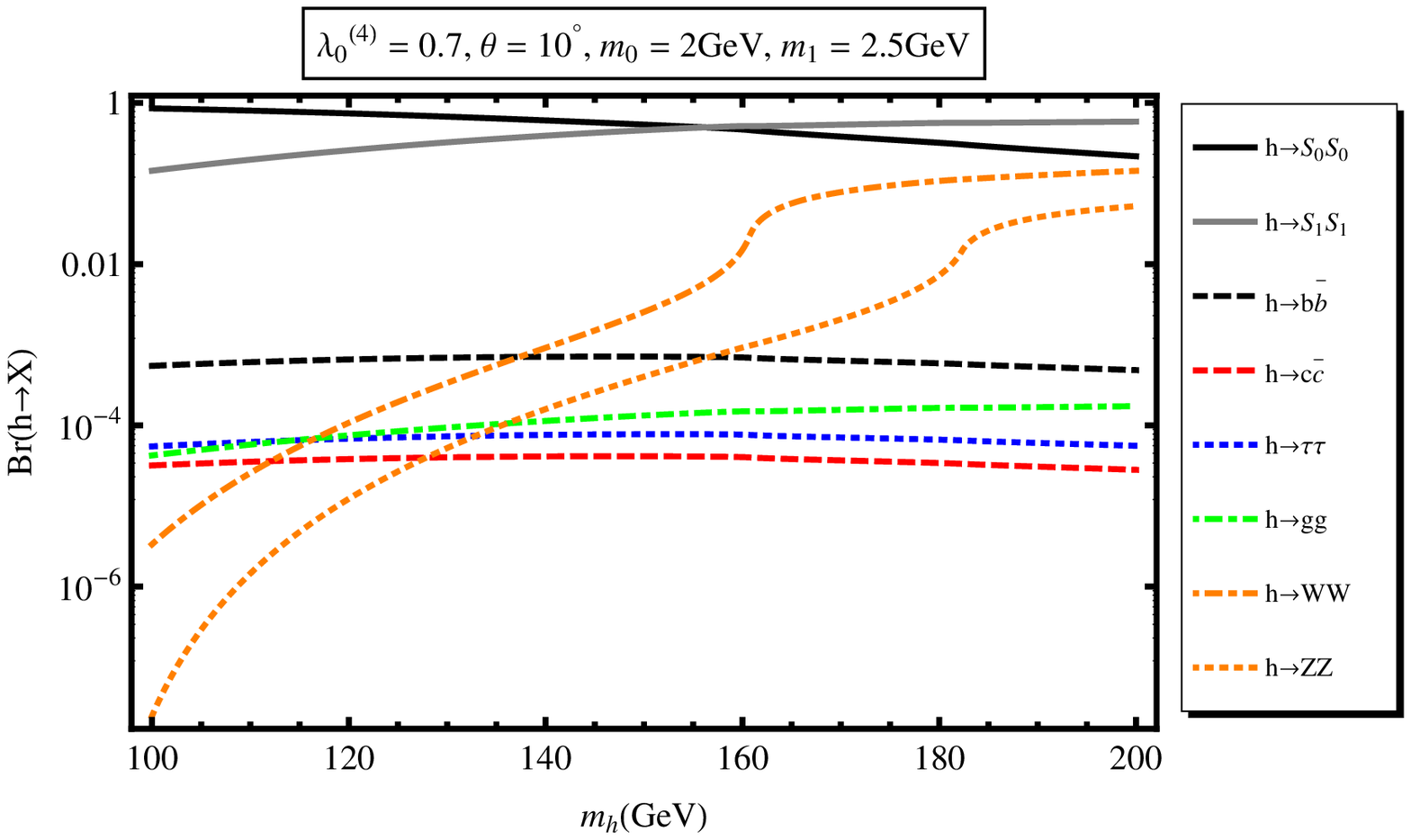}
\caption{Branching ratios for Higgs decays. Larger dark-matter Higgs
coupling.}
\label{log-higgs_lambda04-07_theta-10_m0-2_m1-25_v2}
\end{figure}

However, the decay distribution between $S_{0}$ and $S_{1}$ is not even. The
most dramatic effect comes from the coupling constant $\lambda _{0}^{(4)}$.
When it is very small, the dominant production is that of a pair of $S_{1}$.
This is exhibited in figure \ref
{log-higgs_lambda04-001_theta-10_m0-2_m1-25_v2} for which $\lambda
_{0}^{(4)}=0.01$. As it increases, there is a gradual shift towards a more
dominating dark-matter pair production, a shift competed against by an
increase in $\theta $. Figure \ref
{log-higgs_lambda04-01_theta-10_m0-2_m1-25_v2} displays the branching ratios
for $\lambda _{0}^{(4)}=0.1$ and figure \ref
{log-higgs_lambda04-07_theta-10_m0-2_m1-25_v2} for the larger value $\lambda
_{0}^{(4)}=0.7$. In general, increasing $\theta $ smoothens the crossings of
the $WW$ and $ZZ$ thresholds, and lowers the production of everything except
that of a pair of $S_{1}$, which is instead increased.

Like in the Standard Model, the production of a pair of $b$ quarks dominates
over the production of the other fermions, and all fermions are not favored
by increasing $\lambda _{0}^{(4)}$. Changes in $m_{0}$ and $m_{1}$ have very
little direct effects on all the branching ratios except that of $S_{0}S_{0}$
production where, at small $\theta $, increasing $m_{1}$ ($m_{0}$) increases
(decreases) the branching ratio, with reversed effects at larger $\theta $.
Note though that these masses have indirect impact through the relic density
constraint by excluding certain regions \cite{abada-ghaffor-nasri}.

\section{Concluding remarks}

In this work, we have explored some phenomenological aspects of a
two-singlet extension of the Standard Model we proposed as a simple model
for light cold dark matter. We have looked into the rare decays of $\Upsilon 
$ and $B$ mesons and studied the implications of the model on the decay
channels of the Higgs particle.

For both $\Upsilon $ and $B$ decays, apart from combining with the other two
parameters in the relic-density and perturbativity constraints to exclude
regions of applicability of the model, the Higgs-dark-matter coupling
constant $\lambda _{0}^{(4)}$ and the dark-matter mass $m_{0}$ have little
effect on the shapes of the branching ratios. Also, the effect of increasing
the $h-S_{1}$ mixing angle $\theta $ is to enhance all branching ratios. For 
$\Upsilon $ decays, the dark-matter channel, when kinematically allowed ($
m_{1}\geq 2m_{0}$), dominates over the other decay modes. It reaches the
experimental invisible upper bound for already fairly small values of $
\theta $ and $m_{0}$. From $B^{+}$ decays, we learn that our model is
excluded for $m_{1}<4.8\mathrm{GeV}$ ($=m_{B}-m_{K}$) and $m_{0}<m_{1}/2$.
From $B_{s}$ decay into muons, we learn that for the model to contribute a
distinct signal to this process, it is best to restrict $4\mathrm{GeV}
\lesssim m_{1}\lesssim 6.5\mathrm{GeV}$ with no additional constraint on $
m_{0}$. Also, in general, keeping $\lambda _{0}^{(4)}\lesssim 0.1$ to avoid
systematic exclusion from direct detection for all these processes is safe.

Before closing this section, we comment on the effect that dark matter in
our model has on Higgs searches. Since $m_{h}\gg 2m_{0}$, the process $
h\rightarrow S_{0}S_{0}$ is kinematically allowed and, for a large range of
the parameter space, the ratio 
\begin{equation}
\mathcal{R}_{\mathrm{decay}}^{(b)}=\frac{\mathrm{Br}\left( h\rightarrow
S_{0}S_{0}\right) }{\mathrm{Br}(h\rightarrow b\bar{b})}
\label{ratio-Br-hS0S0-hbb}
\end{equation}
can be larger than one for $m_{h}<120\mathrm{GeV}$ as can be seen in figure 
\ref{log-higgs_lambda04-01_theta-10_m0-2_m1-25_v2}. In this situation, the
LEP bound on the Higgs mass can be weaker. Also, in our model, the Higgs
production at LEP via Higgstrahlung can be smaller than the one in the
Standard Model, and so the Higgs can be as light as $100$GeV. Such a light
Higgs would be in good agreement with the electroweak precision tests\textbf{
\ .}

As to the Higgs searches at the LHC, the ATLAS and CMS collaborations have
reported recently the exclusion of a Higgs mass in the interval 145 -- 460
GeV \cite{ATLAS, CMS}, which seems to suggest that we should have limited
our analysis of the Higgs branching ratios to $m_{h}<145\mathrm{GeV}$.
However, it is important to note that these experimental constraints apply
to the SM Higgs and can not therefore be used as such if the Higgs
interactions are modified. In our model, the mixing of $h$ with $S_{1}$ will
result in a reduction of the statistical significance of the Higgs discovery
at the LHC. Indeed, the relevant quantity that allows one to use the
experimental limits on Higgs searches to derive constraints on the
parameters of the model is the ratio: 
\begin{equation}
\mathcal{R}_{X_{\mathrm{SM}}}\equiv \frac{\sigma \left( gg\rightarrow
h\right) \mathrm{Br}\left( h\rightarrow X_{\mathrm{SM}}\right) }{\sigma ^{( 
\mathrm{SM})}\left( gg\rightarrow h\right) \mathrm{Br}^{\mathrm{SM}}\left(
h\rightarrow X_{\mathrm{SM}}\right) }=\frac{\cos ^{4}{\theta }}{\cos ^{2}{\
\theta }+\Gamma \left( h\rightarrow X_{\mathrm{inv}}\right) /\Gamma _{h}^{ 
\mathrm{SM}}}.  \label{ratioXSM}
\end{equation}
In this expression, $X_{\mathrm{SM}}$ corresponds to all the Standard Model
particles, $X_{i\mathrm{nv}}={S_{0}S_{0}}${\ and }${S_{1}S_{1}}$, $\sigma $
is a cross-section, $\mathrm{Br}^{\mathrm{SM}}\left( h\rightarrow X\right) $
the branching fraction of the SM Higgs decaying into any kinematically
allowed mode $X$, and $\Gamma _{h}^{\mathrm{SM}}$ the total Higgs decay rate
in the Standard Model. To open up the region $m_{h}>140\mathrm{GeV}$
requires the ratio $\mathcal{R}_{X_{\mathrm{SM}}}$ to be smaller than $0.25$ 
\cite{ATLAS, CMS}, a constraint easily fulfilled in our model. By
comparison, the minimal extensions of the Standard Model with just one
singlet scalar or a Majorana fermion, even under $Z_{2}$ symmetry, are
highly constrained in this regard \cite{DarkSide}.

Beyond the present investigation, many other aspects of the model, partly in
relation with the current LHC searches, are just waiting to be explored.

\end{document}